\documentclass[copyright,creativecommons]{eptcs}
\providecommand{\event}{LFMTP 2019} 
\usepackage{breakurl}             
\usepackage{underscore}           

\usepackage{proof}
\usepackage{bussproofs}
\usepackage{tabularx}
\usepackage{alltt}
\usepackage{amssymb}
\usepackage{amsmath}
\usepackage{xcolor}

\newcommand{\E}[2]{\ensuremath{{\epsilon}}}

\newcommand{\LF}  {\mbox{$\mathsf {LF}$}}

\newcommand{\LLFP}  {\mbox{$\mathsf{LLF}_{\mathcal P}$}}

\newcommand{\CLLFPQ} {\mbox{$\mathsf{CLLF}_{\mathcal P?}$}}

\newcommand {\eg}        {{\textit{e}.\textit{g}.}}

\newcommand {\ie}        {{\textit{i}.\textit{e}.}}

\newcommand {\wrt}       {{\textrm{w}.\textrm{r}.\textrm{t}.}}

\newcommand {\at}   {\,}               
\newcommand {\of}   {{:}}              
\newcommand {\Type} {{\mathsf {Type}}} 
\newcommand {\sig}  {{\mathsf {sig}}}  

\newcommand {\Dom} {{\mathsf {Dom}}} 

\newcommand {\App}    [2] {{#1} \at {#2}}                         
\newcommand {\Prod}   [3] {\Pi {#1} \of {#2}.{#3}}            
\newcommand {\Abs}    [3] {\lambda {#1} \of {#2}.{#3}}   

\newcommand {\Lock}   [4] {{\mathcal{L}}^{#1}_{#2, #3}[{#4}]} 

\newcommand {\Unlock} [4] {{\mathcal{U}}^{#1}_{#2, #3}[{#4}]} 

\newcommand {\VDASH}  {\vdash}
\newcommand {\VDASHS} {\vdash_\Sigma} 
\newcommand {\VDASHO} {\vdash_\Omega} 

\renewcommand {\P} {\mathcal{P}} 

\renewcommand {\L} {\mathcal{L}} 
\newcommand   {\U} {\mathcal{U}} 

\newcommand {\eqBL} {{=}_{\!\beta\mathcal{L}}} 

\newcommand {\eval}   {\rightarrow}       
\newcommand {\evalBL} {\eval_{\!\beta\L}} 

\newcommand {\multieval}[1] {\mathop{\eval\!\!\!\!\!\eval_{#1}}} 
\newcommand {\multievalBL}  {\multieval{\!\beta\mathcal{L}}}     

\newcommand{\rew}[1]  {\hspace{-#1mm}}

\newcommand{\ROL}	{(O{\cdot}Lock)}		

\newcommand{\BOMain}	{(\beta{\cdot}O{\cdot}Main)}	
\newcommand{\LOMain}	{(\L{\cdot}O{\cdot}Main)}	

\newcommand{\BLEqMain}  {(\beta\L{\cdot}Eq{\cdot}Main)}
\newcommand{\BLEqRefl}  {(\beta\L{\cdot}Eq{\cdot}Refl)}
\newcommand{\BLEqSym}   {(\beta\L{\cdot}Eq{\cdot}Sym)}
\newcommand{\BLEqTrans} {(\beta\L{\cdot}Eq{\cdot}Trans)}

\newcommand{\CCKPa} {(K{\cdot}\Pi_1{\cdot}\beta\mathcal{L})}
\newcommand{\CCKPb} {(K{\cdot}\Pi_2{\cdot}\beta\mathcal{L})}

\newcommand{\CCFPa} {(F{\cdot}\Pi_1{\cdot}\beta\mathcal{L})}
\newcommand{\CCFPb} {(F{\cdot}\Pi_2{\cdot}\beta\mathcal{L})}
\newcommand{\CCFAa} {(F{\cdot}A_1{\cdot}\beta\mathcal{L})}
\newcommand{\CCFAb} {(F{\cdot}A_2{\cdot}\beta\mathcal{L})}
\newcommand{\CCFLa} {(F{\cdot}\L_1{\cdot}\beta\mathcal{L})}
\newcommand{\CCFLb} {(F{\cdot}\L_2{\cdot}\beta\mathcal{L})}
\newcommand{\CCFLc} {(F{\cdot}\L_3{\cdot}\beta\mathcal{L})}

\newcommand{\CCOAba} {(O{\cdot}\lambda_1{\cdot}\beta\mathcal{L})}
\newcommand{\CCOAbb} {(O{\cdot}\lambda_2{\cdot}\beta\mathcal{L})}
\newcommand{\CCOApa} {(O{\cdot}A_1{\cdot}\beta\mathcal{L})}
\newcommand{\CCOApb} {(O{\cdot}A_2{\cdot}\beta\mathcal{L})}
\newcommand{\CCOLa}  {(O{\cdot}\L_1{\cdot}\beta\mathcal{L})}
\newcommand{\CCOLb}  {(O{\cdot}\L_2{\cdot}\beta\mathcal{L})}
\newcommand{\CCOLc}  {(O{\cdot}\L_3{\cdot}\beta\mathcal{L})}
\newcommand{\CCOUa}  {(O{\cdot}\U_1{\cdot}\beta\mathcal{L})}
\newcommand{\CCOUb}  {(O{\cdot}\U_1{\cdot}\beta\mathcal{L})}
\newcommand{\CCOUc}  {(O{\cdot}\U_1{\cdot}\beta\mathcal{L})}

\def \LF	{\mbox {{\sf LF}}}

\newtheorem{definition}{Definition}

\newcommand{\vdashv} {\vdash_{\Sigma_v}}
\newcommand{\rangeiN} {{\iota \in [1..m]}}
\newcommand{\rangeiNinf} {{\iota \in [0..\infty]}}
\newcommand{\loads} {\to}

\newcommand{\Coq}{\texttt{Coq}}

\title{A Definitional Implementation of the Lax Logical Framework \LLFP\ in \Coq, for Supporting Fast and Loose Reasoning}

\author{Fabio Alessi \quad
Alberto Ciaffaglione \quad
Pietro Di Gianantonio \quad
Furio Honsell \quad
Marina Lenisa
\institute{Department of Mathematics, Computer Science and Physics\\
University of Udine\thanks{Work supported by the Italian departmental research project ``LambdaBridge'' (D.R.N. 37 427/2018 of 03/08/2018, University of Udine).}\\
Udine, Italy}
\email{name.surname@uniud.it}
}

\def\titlerunning{\LLFP\ for Fast \&\ Loose Reasoning}
\def\authorrunning{Fabio Alessi et al.}

\begin{document}
\maketitle

\begin{abstract}
The Lax Logical Framework, \LLFP, was introduced, by a team including the last two authors, to provide a conceptual framework for integrating different proof development tools, thus allowing for {\em external evidence} and for {\em postponing, delegating}, or {\em factoring-out}  side conditions. In particular, \LLFP\  allows for  {\em reducing} the number of times a {\em proof-irrelevant} check is performed. In this paper we give a shallow, actually {\em definitional}, implementation of \LLFP\ in \Coq, \ie\ we use \Coq\ both as  host framework and  oracle for \LLFP. This illuminates the principles underpinning the mechanism of {\em Lock-types} and also suggests how to possibly extend \Coq\ with the features of \LLFP. The derived proof editor is then put to use for developing case-studies on an emerging paradigm, both at logical and implementation level, which we call {\em fast and loose reasoning} following Danielsson et {\em alii} \cite{FL}. This  paradigm trades off efficiency for correctness and amounts to postponing, or running in parallel, tedious or computationally demanding checks, until we are really sure that the intended goal can be achieved. Typical examples are branch-prediction in CPUs and optimistic concurrency control.
\end{abstract}

\section{Introduction}\label{sec:intro}

The \emph{Lax Logical Framework} \LLFP\ is a conservative extension of \LF. It was introduced in \cite{lmcs:3771} with the goal of {\em factoring-out, postponing},  or {\em delegating} to  external tools the verification of those time-consuming judgments, which are ``morally'' {\em proof-irrelevant}. This system was the final step of a series of papers stemming from \cite{HHP-92,HLL06}, aiming at integrating different sources of evidence in a unique Logical Framework. Evidence may derive more conveniently, in effect, from special-purpose external proof search tools, external oracles, or even alternative, non-apodictic, epistemic sources, \eg\ explicit computations according to the Poincar\'e Principle \cite{bar02}, diagrams, or just physical analogies.
The $\Lock{\P}{N}{\sigma}{\cdot}$ constructor was introduced as the appropriate type constructor for expressing {\em inhabitability up-to}. It turned out  to be smoothly expressible as a monad, see \cite{lmcs:3771}, for details.

In this paper, we capitalize in particular on that feature of \LLFP\ which allows for {\em postponing} the checking of  {\em proof-irrelevant} side-conditions, in order to streamline formal reasoning according to an emerging paradigm both at logical and at implementation level. We call this paradigm,  ``fast and loose reasoning'', following \cite{FL}. This paradigm trades off efficiency for correctness and amounts to postponing, or running in parallel, tedious or computationally demanding checks, until we are really sure that the intended goal can be achieved. At logical level this paradigm amounts to the ordinary practice in everyday mathematics based on {\em n\"aive} Set Theory or in programming, based on conjecturing and introducing blanket assumptions, to be checked or formalized later, see \eg\ \cite{CSW14,APLAS}. At the level of implementations natural examples of this paradigm occur both in {\em computer architecture} and {\em concurrency control}, \eg\ {\em branch prediction} in CPUs and {\em optimistic concurrency} in distributed systems \cite{KR1981}. In both cases efficiency is improved by  ``forgetting'', \ie\ running in parallel, time-demanding tests which otherwise would significantly slow down the computation, if carried out sequentially. Of course in the event that the outcome of the test is negative there might be an extra cost for backtracking and restoring the original context. But the trade-off in speed  when this does not occur compensates significantly this drawback.

The case studies in \LLFP, carried out in this paper, namely \emph{call-by-value $\lambda$-calculus} and {\em branch prediction} for URM machines (see \cite{Cutland:computability}) suggest natural extensions of \LLFP\ itself, for expressing nested lock-types. This was already envisaged in \cite{lmcs:3771}. Furthermore, when the predicate in the lock-type is decidable, the case-study on branch prediction suggests to consider the encoding of alternatives as a sort of {\em sum type}. We briefly sketch how to generalize these extensions of \LLFP\ to a full algebra of predicates.

In order to prototype quickly an implementation of \LLFP\ which supports  mechanized proof search, we implement a {\em shallow} encoding of \LLFP\ in the \Coq\ proof assistant. ``Shallow'' in this context means that we delegate as much as possible the mechanics of \LLFP\ to the metalanguage of the host system. Actually the lock-types are rendered by a \Coq\ {\tt Definition}. This is quite interesting in itself, both in exposing the principles underpinning lock-types as well as the bearing it has on proving that predicates are {\em well-behaved}, and conversely, by suggesting how to extend \Coq\ with a Lock constructor.

The authors express their gratitude to Dr. Ivan Scagnetto for many inspiring discussions and comments. They also thank the anonymous referees for their helpful suggestions.

In Section \ref{sec:LLFP} we recap \LLFP. In Section \ref{sec:implementation} we give the implementation in \Coq.
In the two following sections we briefly outline in \LLFP\ paradigmatic applications: call-by-value $\lambda$-calculus and branch prediction for URM machines \cite{Cutland:computability}.
In Section \ref{sec:LLFP-ext} we outline possible extensions of the Lock constructor.
We briefly discuss future directions in Section \ref{sec:future}.
The web appendix of the paper is online at \cite{web-appendix}.

\section{The \LLFP\ logical framework}\label{sec:LLFP}

In this section, following the standard pattern and conventions of
\cite{HHP-92}, we introduce the syntax and the rules of \LLFP, see \cite{lmcs:3771} for more details.
In Figure~\ref{fig:lfpsyntax}, we give the syntactic categories of \LLFP,
namely signatures, contexts, kinds, families (\ie\ types) and objects
(\ie\ terms).
The language is essentially that of classical \LF\ \cite{HHP-92}, to which we add the \emph{lock types} constructor ($\mathcal L$) for building types of the shape $\Lock{\P}{N}{\sigma}{\rho}$, where $\P$ is a predicate on typed judgments. Correspondingly, at the object level, we introduce the lock \emph{constructor} ($\mathcal L$) and the
unlock \emph{destructor} ($\mathcal U$).  The intended meaning of the
$\Lock{\P}{N}{\sigma}{\cdot}$ constructor is that of a \emph{logical
	  filter} expressing inhabitability ``up-to" the verification of $\P(N {:} \sigma)$.

The rules for the main one-step $\beta\L$-reduction, which combines the standard $\beta$-reduction with the novel $\L$-reduction (behaving as a
lock-releasing mechanism, erasing the $\U$-$\L$ pair in a term of the
form $\Unlock {\P} N \sigma {\Lock {\P} N \sigma M}$) appear in Figure~\ref{fig:mainred}.
The rules for one-step closure under context for kinds, families, objects are collected in Figures~\ref{fig:cuckinds},~\ref{fig:blcc},~\ref{fig:cucobjects}, respectively.
We denote the reflexive and transitive closure of $\evalBL$ by
$\multievalBL$. Hence, $\beta\L$-definitional equality is defined in
the standard way, as the reflexive, symmetric, and transitive closure
of $\beta\L$-reduction on kinds, families, objects, as illustrated
in Figure~\ref{fig:bldefeq}. 

\begin{figure}[t!]
\begin{normalsize}
  \begin{center}
      $
      \begin{array}{rcl@{\quad}rcl}
        \Sigma & \in & \emph{Signatures} &
        \Sigma & ::= & \emptyset \mid \Sigma,a \of K \mid \Sigma, c \of \sigma
        \\[1mm]

        \Gamma & \in & \emph{Contexts} &
        \Gamma & ::= & \emptyset \mid \Gamma, x \of \sigma
        \\[1mm]

        K & \in &  \emph{Kinds}  & K & ::= &
        \Type \mid {\Prod x \sigma K}
        \\[1mm]

        \sigma,\tau, \rho & \in & \emph{Families (Types)}
        & \sigma & ::= & a \mid {\Prod x \sigma \tau} \mid {\App \sigma N} \mid
        {\Lock {\P} N \sigma \rho}
        \\[1mm]

        M, N & \in & \emph{Objects} &
        M & ::= &
        c \mid x \mid {\Abs x \sigma M} \mid {\App M N} \mid
        \Lock {\P} N \sigma M \mid {\Unlock {\P} N \sigma M}
      \end{array}
      $
    \end{center}
\vspace{-3mm}
  \caption{The pseudo-syntax of \LLFP}
  \label{fig:lfpsyntax}
\end{normalsize}
\end{figure}

\begin{figure}[t!]
\begin{normalsize}
\begin{center}
  $({\Abs x \sigma M}) \at N \evalBL M[N/x]$ \
  $\BOMain$
\qquad\quad
  $\Unlock {\P} N \sigma {\Lock {\P} N \sigma M} \evalBL M$ \
  $\LOMain$
  \caption{Main one-step-$\beta\mathcal{L}$-reduction rules}
  \label{fig:mainred}
\end{center}
\end{normalsize}
\end{figure}

\begin{figure}[!h]
	\begin{center}
		\begin{tabularx}{16cm}{m{5cm}m{2cm}m{5cm}m{2cm}}
			\begin{prooftree}
				\AxiomC{$\sigma \evalBL \sigma'$}
				\UnaryInfC{${\Prod x \sigma K} \evalBL {\Prod x {\sigma'} K}$}
			\end{prooftree} & $\CCKPa$ &
			\begin{prooftree}
				\AxiomC{$K \evalBL K'$}
				\UnaryInfC{${\Prod x \sigma K} \evalBL {\Prod x \sigma {K'}}$}
			\end{prooftree} & $\CCKPb$ \\
		\end{tabularx}
  \end{center}
\vspace{-3mm}
  \caption{$\beta\mathcal{L}$-closure-under-context for kinds}
  \label{fig:cuckinds}
\end{figure}

\begin{figure}[!h]
  \begin{center}
    \begin{tabularx}{16cm}{m{5cm}m{2cm}m{5cm}m{2cm}}
      \begin{prooftree}
        \AxiomC{$\sigma \evalBL \sigma'$}
        \UnaryInfC{${\Prod x \sigma \tau} \evalBL {\Prod x {\sigma'} \tau}$}
      \end{prooftree} & $\CCFPa$ &
      \begin{prooftree}
        \AxiomC{$\tau \evalBL \tau'$}
        \UnaryInfC{${\Prod x \sigma \tau} \evalBL {\Prod x \sigma {\tau'}}$}
      \end{prooftree} & $\CCFPb$
\\
      \begin{prooftree}
        \AxiomC{$\sigma \evalBL \sigma'$}
        \UnaryInfC{${\App \sigma N} \evalBL {\App {\sigma'} N}$}
      \end{prooftree} & $\CCFAa$ &
      \begin{prooftree}
        \AxiomC{$N \evalBL N'$}
        \UnaryInfC{${\App \sigma N} \evalBL {\App \sigma {N'}}$}
      \end{prooftree} & $\CCFAb$
\\
      \begin{prooftree}
        \AxiomC{$N \evalBL N'$}
        \UnaryInfC{${\Lock {\P} N \sigma \rho} \evalBL {\Lock {\P} {N'} \sigma \rho}$}
      \end{prooftree} & $\CCFLa$ &
      \begin{prooftree}
        \AxiomC{$\sigma \evalBL \sigma'$}
        \UnaryInfC{${\Lock {\P} N \sigma \rho} \evalBL {\Lock {\P} N {\sigma'} \rho}$}
      \end{prooftree} & $\CCFLb$
\\
      \begin{prooftree}
        \AxiomC{$\rho \evalBL \rho'$}
        \UnaryInfC{${\Lock {\P} N \sigma \rho} \evalBL {\Lock {\P} N \sigma {\rho'}}$}
      \end{prooftree} & $\CCFLc$ & &
\\
    \end{tabularx}
  \end{center}
\vspace{-3mm}
  \caption{$\beta\mathcal{L}$-closure-under-context for families}
  \label{fig:blcc}
\end{figure}

\begin{figure}[!h]
	\begin{center}
		\begin{tabularx}{16cm}{m{5cm}m{2cm}m{5cm}m{2cm}}
			\begin{prooftree}
      	\AxiomC{$\sigma \evalBL \sigma'$}
      	\UnaryInfC{${\Abs x \sigma M} \evalBL {\Abs x {\sigma'} M}$}
			\end{prooftree} & $\CCOAba$ &
        \begin{prooftree}
 		     \AxiomC{$M \evalBL M'$}
 		     \UnaryInfC{${\Abs x \sigma M} \evalBL {\Abs x \sigma {M'}}$}
			\end{prooftree} & $\CCOAbb$ \\
			\begin{prooftree}
  	   		 \AxiomC{$M \evalBL M'$}
  			    \UnaryInfC{${\App M N} \evalBL {\App {M'} N}$}
			\end{prooftree} & $\CCOApa$ &
			\begin{prooftree}
  		    \AxiomC{$N \evalBL N'$}
 		    \UnaryInfC{${\App M N} \evalBL {\App M {N'}}$}
			\end{prooftree} & $\CCOApb$ \\
			\begin{prooftree}
      		\AxiomC{$N \evalBL N'$}
      		\UnaryInfC{${\Lock {\P} N \sigma M} \evalBL {\Lock {\P} {N'} \sigma M}$}
			\end{prooftree} & $\CCOLa$ &
			\begin{prooftree}
   		   \AxiomC{$\sigma \evalBL \sigma'$}
   		   \UnaryInfC{${\Lock {\P} N \sigma M} \evalBL {\Lock {\P} N {\sigma'} M}$}
			\end{prooftree} & $\CCOLb$ \\
			\begin{prooftree}
      		\AxiomC{$M \evalBL M'$}
      		\UnaryInfC{${\Lock {\P} N \sigma M} \evalBL {\Lock {\P} N \sigma {M'}}$}
			\end{prooftree} & $\CCOLc$ &
			\begin{prooftree}
      		\AxiomC{$N \evalBL N'$}
      		\UnaryInfC{${\Unlock {\P} N \sigma M} \evalBL {\Unlock {\P} {N'} \sigma M}$}
			\end{prooftree} & $\CCOUa$ \\
			\begin{prooftree}
   		   \AxiomC{$\sigma \evalBL \sigma'$}
   		   \UnaryInfC{${\Unlock {\P} N \sigma M} \evalBL {\Unlock {\P} N {\sigma'} M}$}
			\end{prooftree} & $\CCOUb$ &
			\begin{prooftree}
      		\AxiomC{$M \evalBL M'$}
      		\UnaryInfC{${\Unlock {\P} N \sigma M} \evalBL {\Unlock {\P} N \sigma {M'}}$}
			\end{prooftree} & $\CCOUc$ \\
		\end{tabularx}
  \end{center}
  \caption{$\beta\mathcal{L}$-closure-under-context for objects}
  \label{fig:cucobjects}
\end{figure}

\begin{figure}[!h]
  \begin{center}
    \begin{tabularx}{16cm}{m{3cm}m{2cm}m{5cm}m{2cm}}
			\begin{prooftree}
      	\AxiomC{$T \evalBL T'$}
      	\UnaryInfC{$T \eqBL T'$}
			\end{prooftree} & $\BLEqMain$ &
			\begin{prooftree}
 		     \AxiomC{}
 		     \UnaryInfC{$T \eqBL T$}
			\end{prooftree} & $\BLEqRefl$ \\
			\begin{prooftree}
  	   		 \AxiomC{$T \eqBL T'$}
  			    \UnaryInfC{$T' \eqBL T$}
			\end{prooftree} & $\BLEqSym$ &
			\begin{prooftree}
  		    \AxiomC{$T  \eqBL T'$}
  		    \AxiomC{$T' \eqBL T''$}
 		    \BinaryInfC{$T \eqBL T''$}
			\end{prooftree} & $\BLEqTrans$ \\
    \end{tabularx}	
  \end{center}
  \caption{$\beta\L$-definitional equality}
  \label{fig:bldefeq}
\end{figure}
\begin{figure}[t!]
  \begin{normalsize}
    \begin{center}
      $
      \begin{array}[t]{l}

        \mbox{\sf Signature~rules}\hfill
        \\[2mm]

        \infer[(S{\cdot}Empty)]
        {\emptyset\ \sig}
        {}
        \\[3mm]

        \infer[(S{\cdot}Kind)]
        {\Sigma, a \of K\ \sig}
        {\begin{array}{l@{\quad}l}
            \VDASHS K & a \not \in \Dom(\Sigma)
          \end{array}}
        \\[3mm]

        \infer[(S{\cdot}T\!ype)]
        {\Sigma, c \of \sigma\ \sig}
        {\begin{array}{l@{\quad}l}
            \VDASHS \sigma \of \Type & c \not \in \Dom(\Sigma)
          \end{array}}
        \\[3mm]

        \mbox{\sf Context~rules}\hfill
        \\[2mm]

        \infer[(C{\cdot}Empty)]
        {\VDASHS \emptyset}
        {\Sigma\ \sig}
        \\[3mm]

        \infer[(C{\cdot}T\!ype)]
        {\VDASHS \Gamma, x \of \sigma}
        {\begin{array}{l@{\quad}l}
            \Gamma \VDASHS \sigma \of \Type & x \not \in \Dom(\Gamma)
          \end{array}}
        \\[3mm]

        \mbox{\sf Kind~rules}\hfill
        \\[2mm]

        \infer[(K{\cdot}T\!ype)]
        {\Gamma \VDASHS \Type}
        {\VDASHS \Gamma}
        \\[3mm]

        \infer[(K{\cdot}Pi)]
        {\Gamma \VDASHS {\Prod x \sigma K}}
        {\Gamma, x \of \sigma \VDASHS K}
        \\[4mm]

        \mbox{\sf Family~rules}\hfill
        \\[2mm]

        \infer[(F{\cdot}Const)]
        {\Gamma \VDASHS a : K}
        {\VDASHS \Gamma & a \of K \in \Sigma}
        \\[2mm]

        \infer[(F{\cdot}Pi)]
        {\Gamma \VDASHS {\Prod x \sigma \tau} : \Type}
        {\Gamma, x \of \sigma \VDASHS \tau : \Type}
        \\[-6mm]

      \end{array}
      \rew{67}
      \begin{array}[t]{r}
        \\[1mm]

        \infer[(F{\cdot}App)]
        {\Gamma \VDASHS {\App \sigma N} : K[N/x]}
        {\Gamma \VDASHS \sigma : {\Prod x \tau K} & \Gamma \VDASHS N : \tau}
        \\[3mm]

        \infer[(F{\cdot}Lock)]
        {\Gamma \VDASHS {\Lock {\P} N \sigma {\rho}} : \Type}
        {\Gamma \VDASHS \rho : \Type & \Gamma \VDASHS N : \sigma}
        \\[3mm]

        \infer[(F{\cdot}Conv)]
        {\Gamma \VDASHS \sigma : K'}
        {\Gamma \VDASHS \sigma : K &
          \Gamma \VDASHS K' & K \eqBL K'}
        \\[3mm]

        \mbox{\sf Object~rules}\hfill
        \\[2mm]

        \infer[(O{\cdot}Const)]
        {\Gamma \VDASHS c : \sigma}
        {\VDASHS \Gamma & c \of \sigma \in \Sigma}
        \\[3mm]

        \infer[(O{\cdot}V\!ar)]
        {\Gamma \VDASHS x : \sigma}
        {\VDASHS \Gamma & x \of \sigma \in \Gamma}
        \\[3mm]

        \infer[(O{\cdot}Abs)]
        {\Gamma \VDASHS {\Abs x \sigma M} : {\Prod x \sigma \tau}}
        {\Gamma, x \of \sigma \VDASHS M : \tau}
        \\[3mm]

        \infer[(O{\cdot}App)]
        {\Gamma \VDASHS M \at N : \tau[N/x]}
        {\Gamma \VDASHS M : {\Prod x \sigma \tau} & \Gamma \VDASHS N : \sigma}
        \\[3mm]

        \infer[(O{\cdot}Conv)]
        {\Gamma \VDASHS M : \tau}
        {
          \Gamma \VDASHS M : \sigma & 
          \Gamma \VDASHS \tau : \Type & \sigma \eqBL \tau
        }
        \\[3mm]

        \infer[(O{\cdot}Lock)]
        {\Gamma \VDASHS {\Lock {\P} N \sigma M} : {\Lock {\P} N \sigma {\rho}}}
        {\Gamma \VDASHS M : \rho & \Gamma \VDASHS N : \sigma}
        \\[3mm]

        \infer[(O{\cdot}Top{\cdot}Unlock)]
        {\Gamma \VDASHS {\Unlock {\P} N \sigma M} : \rho}
        {\Gamma \VDASHS M : {\Lock {\P} N \sigma \rho}  &
          \P(\Gamma \VDASHS N : \sigma)}
        \\[3mm]

        \infer[(F{\cdot}Guarded{\cdot}Unlock)]
    {\Gamma  \VDASHS \Lock {\P} S \sigma { \rho[\Unlock {\P} {S'} {\sigma'} N /x]} : \Type}
    {\Gamma, x\of\tau \VDASHS \Lock {\P} S \sigma {\rho} : \Type & \Gamma \VDASHS N :  \Lock {\P} {S'} {\sigma'} \tau & \sigma \eqBL \sigma' & S \eqBL S' } 
            \\[3mm]

            \infer[(O{\cdot}Guarded{\cdot}Unlock)]
            {\Gamma  \VDASHS {\Lock {\P} S \sigma {M[\Unlock {\P} {S'}
            {\sigma'}{N} /x]}} :
            \Lock {\P} S \sigma { \rho[\Unlock {\P} {S'} {\sigma'} {N} /x]}}
            {\Gamma, x\of\tau \VDASHS \Lock {\P} S \sigma {M} : \Lock {\P} S \sigma {\rho} &
                  \Gamma \VDASHS N :  \Lock {\P} {S'} {\sigma'} {\tau} &
                                                                         \sigma \eqBL \sigma' & S \eqBL S' }
      \end{array}
      $
    \end{center}
  \end{normalsize}
  \caption{The \LLFP\ Type System}
  \label{fig:lfptypesys}
\end{figure}
Following the standard specification paradigm of Constructive Type Theory, we define lock-types using \emph{introduction}, \emph{elimination}, and \emph{equality rules}. Namely, see Figure~\ref{fig:lfptypesys}, we introduce a lock-\emph{constructor} for building objects
$\Lock{\P}{N}{\sigma}{M}$
of type $\Lock{\P}{N}{\sigma}{\rho}$,
via the \emph{introduction rule} $\ROL$.
Correspondingly, we introduce an unlock-\emph{destructor}
$\Unlock{\P}{N}{\sigma}{M}$
via the \emph{elimination rule} $(O{\cdot} Guarded{\cdot} Unlock)$, which is reminiscent in its shape of a Gentzen-style {\em left-introduction} rule.
In order to provide the intended meaning of
$\Lock{\P}{N}{\sigma}{\cdot}$, we need to introduce in \LLFP\ also the
rule $(O{\cdot}Top{\cdot} Unlock)$, which allows for the elimination
of the lock-type constructor if the predicate $\P$ is verified,
possibly \emph{externally}. Figure~\ref{fig:lfptypesys} shows the full typing system of
\LLFP. All \emph{type equality rules} of \LLFP\ use as notion of conversion $\beta\L$-definitional equality.

One may wonder why the rule $(O{\cdot}Top{\cdot}Unlock)$ is not enough and a $(O{\cdot} Guarded{\cdot} Unlock)$-rule is called for. First of all, releasing a locked term, \ie\ checking a proof-irrelevant side condition is precisely what slows down a derivation. Ultimately we need an external evaluation or to query an external oracle (possibly more than once for the same
property) obtaining a positive answer. Moreover, properties under lock are usually not essential to the main thrust of the proof, because they are {\em proof-irrelevant} and one would like to be free to proceed with the
main argument, postponing, as much as
possible, the verification of ``details''. This is precisely the spirit of the ``fast and loose'' reasoning paradigm \cite{FL}. Namely, when we reach a given stage of a proof development where we are not able, or we do not want to waste time, to verify a side-condition, we may want to \emph{postpone} such a task, unlock immediately the given term, and proceed with
the proof. The $(O{\cdot} Guarded{\cdot} Unlock)$-rule allows us to realize exactly this. The external lock-type of the term within which we release the
unlocked term will preserve safety, keeping track that the verification has to be carried
out at least once, sooner or later.

We conclude this section by recalling that, since external predicates $\P$
affect reductions in \LLFP, they must be \emph{well-behaved} in order
to preserve subject reduction. This property is necessary for achieving
\emph{decidability}, \emph{relative to} an oracle, which is essential
to any proof-checker such as \LLFP.  We introduce, therefore, the following crucial definition, where  $\alpha$ is shorthand for the ``conclusion'' of a judgment.

\begin{definition}[Well-behaved predicates, \cite{honsell:hal-00906391}]
  \label{def:wbred}
  A finite set of predicates $\{ \P_i\}_{i\in I}$ is
  \emph{well-behaved} if each $\P$ in the set satisfies the following
  conditions:
  \begin{enumerate}

  \item {\bf \emph{Closure under signature and context weakening and
        permutation:}}
    \begin{enumerate}
    \item If $\Sigma$ and $\Omega$ are valid signatures such that
      $\Sigma \subseteq \Omega$ and $\P(\Gamma \VDASHS
      \alpha)$, 
      then $\P(\Gamma \VDASHO \alpha)$. 
    \item If $\Gamma$ and $\Delta$ are valid contexts such that
      $\Gamma\subseteq \Delta$ and $\P(\Gamma \VDASHS
      \alpha)$, 
      then \mbox{$\P(\Delta \VDASHS \alpha)$.} 
    \end{enumerate}

  \item{\bf \emph{Closure under substitution:}}\\ If
    $\P(\Gamma, x \of \sigma', \Gamma' \VDASHS N : \sigma)$ and
    $\Gamma \VDASHS N' : \sigma'$, 
    then $\P(\Gamma, \Gamma'[N'/x] \VDASHS N [N'/x] : \sigma[N'/x])$.

  \item{\bf\emph {Closure under reduction:}}
    \begin{enumerate}
    \item If $\P(\Gamma \VDASHS N : \sigma)$ and $N \evalBL N'$,
      then $\P(\Gamma \VDASHS N' : \sigma)$. 
    \item If $\P(\Gamma \VDASHS N : \sigma)$ and
      $\sigma \evalBL \sigma'$, 
      then $\P(\Gamma \VDASHS N :\sigma')$. 
    \end{enumerate}
  \end{enumerate}
\end{definition}

\section{A definitional implementation of \LLFP\ in \Coq}\label{sec:implementation}

An implementation, from scratch, of the logical framework \LLFP\ in a functional language, would definitely be particularly efficient, and has indeed been attempted successfully as far as {\em proof checking}, by Vincent Michielini at ENS Lyon \cite{michielini}. But in order to  provide a rapid prototyping of a full-fledged {\em proof development environment} for \LLFP, we prefer to capitalize on the existing proof-assistant \Coq. This could be done very easily, albeit indirectly, using \Coq\ as a {\em logical metalanguage} by giving an encoding of \LLFP\ in \Coq. But we do not need a ``deep'' encoding of \LLFP's syntactic categories and related judgments, since we are not interested in reasoning on \LLFP's metatheory. Our encoding could be, actually ``should be'', as ``shallow'' as possible so that we may be able to delegate to \Coq 's metalanguage not only all of \LLFP\ metalanguage, but moreover, reduce {\em inhabitation-search} in \LLFP\ to {\em proof-search} in \Coq.

We achieve this by exploiting the fact that \Coq\ is a conservative extension of the dependent constructive type theory of \LF\ \cite{HHP-92} which underpins the type system of \LLFP, \cite{honsell:hal-00906391}. We  simulate/implement, therefore, in \Coq\ the mechanism of lock-types, and use \Coq\ both as the host system and as the oracle for external propositions. This yields a {\em definitional} encoding of \LLFP\ in \Coq. It restricts us, of course, to dealing only with total \Coq-definable predicates, but this is enough for illustrating our approach and moreover has the advantage of enforcing automatically the well-behavedness of the external predicates, provided their \Coq-encoding is adequate. 

In practice, therefore, \LLFP\  \emph{signatures} and \emph{contexts} are not modeled  via structured datatypes, such as \eg\ lists, but are represented by \Coq 's contexts and made available as assumptions.
The \emph{kind} $\Type$ is represented directly via \Coq's sorts {\tt Set} and {\tt Prop}. We will explain below why it is convenient, although not necessary, to use both. Hence \emph{type families}  are rendered as \Coq\ sets or propositions and \emph{objects} as their inhabitants.
Remarkably, we need to implement {\em only} the lock constructor for families, as follows: \begin{verbatim}
Definition lockF := fun s: Set => fun N: s => fun P: s -> Prop =>
                    fun r: Prop => forall x: P N, r.
\end{verbatim}

Families are therefore typed by {\tt Prop} and objects by families, with the exception of the family involved in the definition of the predicate $\P$, which is typed by {\tt Set}. This is what makes possible, in using \Coq\  as the oracle, to take full advantage, in defining the external predicates of \LLFP, of its logical strength in terms of (co)inductive datatypes and (co)recursive functions.

In a nutshell, the gist of the previous definition is to represent the locking of families in \LLFP\ by the  $\Pi$-type:
\[
\ulcorner\Lock{\cal P}{N}{\sigma}{\rho}\urcorner\qquad \leadsto\qquad 
\Pi_{x:{\cal P}(\ulcorner N\urcorner)}\ulcorner \rho\urcorner
\]
This encoding might appear weak, but actually it permits  us to develop formal proofs ``under ${\cal P}$'', just by ``\texttt{unfold}"ing the \texttt{lockF} constructor when it appears in the goal. As a consequence, 
somewhat surprisingly, our \texttt{Definition} is sufficient to derive  {\em all} the typing rules of \LLFP\ that involve lock-types as \Coq's Lemmas:
\begin{itemize}
\item lock-introduction (see rule $(O{\cdot}Lock)$ in Fig.~\ref{fig:lfptypesys}) is rendered by $\Pi$-introduction:
\begin{verbatim}
Lemma lock: forall s: Set, forall N: s, forall P: s -> Prop,
            forall r: Prop, forall M:r, lockF s N P r.
intros; unfold lockF; intro; assumption.
Qed.
\end{verbatim}
\item unlocking at  top level (see rule $(O{\cdot}Top{\cdot}Unlock)$ in Fig.~\ref{fig:lfptypesys}) is rendered by means of $\Pi$-elimination:
\begin{verbatim}
Lemma top_unlock: forall s: Set, forall N: s, forall P: s -> Prop,
                  forall r: Prop, forall M:lockF s N P r, forall x:P N, r.
intros; exact (M x).
Qed.
\end{verbatim}
\item finally, guarded-unlocking (see rule $(O{\cdot}Guarded{\cdot}Unlock)$ in Fig.~\ref{fig:lfptypesys}) {may be rendered in several equivalent ways, which we have experimented with in our work.
In the end, we have chosen to rephrase it in a way where the \texttt{lockF} constructor appears ``\texttt{unfold}"ed in the conclusion of the rule, to support a more flexible management of proofs. Notice, in particular, how the rule is encoded} 
by an interplay of dependencies, namely that  of the unlocked inner term \texttt{(N x)} on the externally bound variable of the outer lock \texttt{x}, and that of the outer locked typed \texttt{(r (N x))} on the unlocked inner term \texttt{(N x)}.
{We will comment further on this rule in the following sections which deal with applications:}
\begin{verbatim}
Lemma guarded_unlock: forall s: Set, forall S: s, forall P: s -> Prop,
                   forall t: Prop, forall r: t -> Prop,
                   forall M: forall y:t, lockF s S P (r y),
                   forall N: lockF s S P t,
                   forall x: P S, r (N x).
intros; unfold lockF; unfold lockF in M; intros; apply M; auto.
Qed.
\end{verbatim}
\end{itemize}

In conclusion we have achieved an encoding of \LLFP\ through a simple {\tt Definition} in \Coq, see \cite{web-appendix}. As pointed out earlier this does not support the full strength of \LLFP, in that predicates are restricted to \Coq-definable terms of some type which eventually maps into {\tt Prop}. Apart from this restriction, however, since \Coq\ is a conservative extension of \LF, the implementation is obviously faithful with respect to all the rules of \LLFP. 

We could give a slightly deeper implementation which, following \cite{lmcs:3771}, would yield a more perspicuous rendering of the monadic nature of Locks.

\section{Call-by-value $\lambda$-calculus}\label{sec:lambda}

In this section we test our implementation of \LLFP\  on a standard benchmark-encoding for Logical Frameworks, namely untyped $\lambda$-calculus with a call-by-value equational theory, \ie\ the $\lambda_v$-calculus.  In the literature there are many ways of encoding this system. We use the signature given  in \cite{lmcs:3771}, because it illustrates the flexibility of \LLFP\ in capitalizing on Higher Order Abstract Syntax (HOAS) when considering {\em bound} variables, while retaining the ordinary way of referring to {\em free} variables. We proceed then to experiment with it in \LLFP\ using the \Coq\ implementation introduced in Section \ref{sec:implementation}.

The well-known abstract syntax of
$ \lambda$-calculus is given by: $M,N ::= x\ |\ M\ N\ |\ \lambda x.M$. We will model {\em free} variables in this
object language as constants in \LLFP.
{\em Bound} variables will be modeled by variables of
the metalanguage, thus  exploiting HOAS in delegating $\alpha$-conversion and capture-avoiding substitution to the metalanguage.
For instance, the $\lambda$-term $x$ (in which the variable is free) is encoded by  the term $\VDASHS${\tt(free n):term} for a suitable (encoding of a) natural number {\tt n} (see Definition~\ref{signature-syntax-ulc} below). On the other hand, the $\lambda$-term $\lambda x.x$ (in which the variable is obviously bound) is encoded by $\VDASHS{\tt(lam}\ \lambda${\tt x:term.x)}. 

We  introduce therefore the following signature:

\begin{definition}[\LLFP\ signature $\Sigma_\lambda$ for untyped $\lambda$-calculus]\label{signature-syntax-ulc}
\[
\begin{array}{lll}
\verb|nat: Type| & \quad
\verb|term: Type| &
\\
\verb|0: nat| & \quad
\verb|S: nat| \to \verb|nat|
\\
\verb|free: nat| \to \verb|term| & \quad
\verb|app: term| \to \verb|term| \to \verb|term| & \quad
\verb|lam: (term| \to \verb|term)| \to \verb|term|
\end{array}
\]
\end{definition}

\noindent We use natural numbers as standard abbreviations for repeated applications of {\tt S} to {\tt 0}. 

\noindent Standard call-by-value conversion is given by the following:

\begin{definition}[Call-by-value {equational theory}]\label{def:cbv}
$$
 \begin{array}{l@{\quad\quad}l}
   \infer[\sf (refl)]
   {\VDASH_{CBV} M=M}
   {}
   &
   \infer[\sf (symm)]
   {\VDASH_{CBV} M=N}
   {\VDASH_{CBV} N=M}
   \\[3mm]
   \infer[\sf (trans)]
   {\VDASH_{CBV} M=P}
   {\VDASH_{CBV} M=N \quad \VDASH_{CBV} N=P}  
   \quad 
   &
   \infer[\sf (app)]
   {\VDASH_{CBV} MM'=NN'}
   {\VDASH_{CBV} M=N \quad \VDASH_{CBV} M'=N'}
   \\[3mm]
   \infer[\sf (\beta_v)]
   {\VDASH_{CBV} (\lambda x.M)v=M[v/x]}
   {\mbox{$v$ is a value}}
   &
   \infer[\sf (\xi_v)]
   {\VDASH_{CBV} \lambda x.M=\lambda x.N}
   {\VDASH_{CBV} M=N}
 \end{array}
$$
where values are either variables or abstractions.
\end{definition}

Accordingly, we extend the signature of Definition \ref{signature-syntax-ulc} as
  follows:

\begin{definition}[\LLFP\ signature $\Sigma_{\sf v}$ for
  $\lambda_{\sf v}$-calculus]\label{signature-reduction-ml}
  
\[
\begin{array}{ll}
\verb|eq:| & \verb|term| \to \verb|term| \to \verb|Type|
\\
\verb|refl:| & \Pi \verb|M:term.| \; \verb|eq M M|
\\
\verb|symm:| & \Pi \verb|M,N:term.| \; \verb|eq M N| \to \verb|eq N M|
\\
\verb|trans:| & \Pi \verb|M,N,P:term.| \; \verb|eq M N| \to \verb|eq N P| \to \verb|eq M P|
\\
\verb|eq_app:| & \Pi \verb|M,N,P,Q:term.| \; \verb|eq M N| \to \verb|eq P Q| \to \verb|eq (app M P) (app N Q)|
\\
\verb|betav:| & \Pi \verb|M:term| \to \verb|term.| \; \Pi \verb|N:term.| \;
	\Lock{\scriptsize\emph{Val}}{\mathtt{N}}{\mathtt{term}}
			{\mathtt{eq\ \,(app\ \,(lam\ \,M)\ \,N)\ \,(M\ \,N)}}
\\
\verb|csiv:| & \Pi \verb|M,N:term| \to \verb|term.| \; 
	\verb|(| \Pi \verb|x:term.| \; 
	\Lock{\scriptsize\emph{Val}}{\mathtt{x}}{\mathtt{term}}
		{\mathtt{eq\ \,(M\ \, x)\ \,(N\ \, x)}} \verb|)| 
		\to {\mathtt{eq\ \, (lam\ \, M)\ \,(lam\ \, N)}}
\end{array}
\]

\noindent where the predicate
{\tt \emph{Val}}$(\Gamma\vdashv$ {\tt N:term}$)$ holds if and only if {\tt N} is either an abstraction or a variable
(\ie\ a term of the shape {\tt{(free i)}}).
\end{definition}
Notice how, in Definition \ref{signature-reduction-ml}, \LLFP's \emph{lock-types} permit us to model the $(\beta_v)$ and $(\xi_v)$ rules: the former holds ``up-to" the verification of {\tt \emph{Val}}$(\Gamma\vdashv$ {\tt N:term}$)$, while the latter depends, in turn, on a locked premise.

We now proceed to represent the above signature in the  \Coq\ editor for \LLFP\ presented in Section \ref{sec:implementation}.
Then we use such a formalization to carry out a simple interactive proof. The full code appears in the on-line appendix, see \cite{web-appendix}.

First, we declare the new kind of terms (typed by {\tt Set}) and their ``constructors", by exploiting the built-in representation of  natural numbers, which lives in {\tt Set}:
\begin{verbatim}
Parameter term: Set.
Parameter free: nat -> term.
Parameter app : term -> term -> term.
Parameter lam : (term -> term) -> term.
\end{verbatim}
Then, we model the predicate {\sf {Val}} in \Coq, since the oracle role is played by the host framework:
\begin{verbatim}
Definition Val := fun N:term => (exists n, N = (free n)) \/
                                (exists M, N = (lam M)).
\end{verbatim}

\noindent One can easily, albeit {\em not formally},  check that the above \Coq-encoding of ``being a value'' is an adequate formalization of the intended concept, thereby giving evidence also, that the predicate originally used in the lock is well-behaved. 

Finally, we encode the call-by-value {equational theory}, by means of a predicate (\ie\ typed by {\tt Prop}):
\begin{verbatim}
Parameter eq: term -> term -> Prop.
Parameter refl:   forall M:term, eq M M.
Parameter symm:   forall M N:term, eq M N -> eq N M.
Parameter trans:  forall M N P:term, eq M N -> eq N P -> eq M P.
Parameter eq_app: forall M N P Q:term, eq M N -> eq P Q -> 
	                  eq (app M P) (app N Q).
Parameter betav:  forall M:term->term, forall N:term,
                  lockF term N Val (eq (app (lam M) N) (M N)).
Parameter csiv:   forall M N:term->term,
                  (forall x:term, lockF term x Val (eq (M x) (N x))) ->
                  eq (lam M) (lam N).
\end{verbatim}
Notice that, in defining {\tt term} and {\tt eq}, we do not use \Coq's inductive types, as these would go beyond \LLFP's expressivity, but we rely on that part of \Coq\ metalanguage which is shared with \LLFP.
We do not use \Coq\ inductive types for encoding terms because we exploit full Higher Order Abstract Syntax (HOAS). We could have used {\em weak} HOAS to deal with variables  but we prefer to stay minimal and avoid exotic terms. 

The use of lock-types in expressing the ($\xi_v$)-rule, although natural, might appear to be unmanageable in applications, since the variable in the premise is not immediately free or bound, but only {\em bindable}. But, as it will become apparent in the following example, the $(O{\cdot}Guarded{\cdot}Unlock)$ rule in \LLFP\ accommodates precisely this issue. Namely, the necessary verification is pushed at the outermost level where it is discharged by the application of the ($\xi_v$)-rule.

To point out the practical value of the \Coq\ editor introduced in this paper, we conclude the section with the formal proof of the simple equation $\lambda x.\ z\ ((\lambda y.y)\ x) = \lambda x.\ z\ x$.
The crucial step is the application of the $(O{\cdot}Guarded{\cdot}Unlock)$ rule: the first premise is given by the application of the $(O{\cdot}Lock)$ rule to the conclusion of the {\tt eq_app} rule, while the second premise is the conclusion of the {\tt betav} rule. Please notice the power of the  $(O{\cdot}Guarded{\cdot}Unlock)$ rule, which allows us to apply the rules of the $\Sigma_{\sf v}$ signature (in this case, the {\tt eq_app} rule), ``under {\sf {Val}}'', \ie\ the latter can handle even premises which are locked
\footnote{Note that in the following proof tree we shorten $(O{\cdot}Guarded{\cdot}Unlock)$ to $(O{\cdot}G{\cdot}U)$ and $(O{\cdot}Lock)$ to $(O{\cdot}L)$ for saving space.}:
\[
\infer[(\textrm{csiv})]
{
z{:}term \vdashv eq(\lambda x{:}term.\ app(z,app(lam (\lambda y {:} term.\, y),x)),
\lambda x{:}term.\ app(z,x))
}
{
\infer[]
{
z{:}term \vdashv \forall x{:}term.\ \Lock{Val}{x}{term}
			{eq(app(z,app(lam (\lambda y {:} term.\, y),x)),app(z,x))}
}
{\infer[\textrm{(eq_app via $O{\cdot}G{\cdot}U$, $O{\cdot}L$})]
		{
		z,x{:}term \vdashv \Lock{Val}{x}{term}
			{eq(app(z,app(lam (\lambda y {:} term.\, y),x)),app(z,x))}
		}
		{\infer[\textrm{(refl)}]
			{
			z{:}term \vdashv eq(z,z)
			}
			{
			}
         &
		\infer[\textrm{(betav)}]
			{
			x{:}term \vdashv \Lock{Val}{x}{term}
				{eq(app(lam (\lambda y {:} term.\, y),x),x)}
			}
			{
			}				
		}
}
}
\]
We conclude by remarking that using the \Coq\ editor of \LLFP\ we may accomplish the above goal without having to exhibit the full proof term beforehand, as we had to in \cite{lmcs:3771}, because we can now build it interactively and incrementally, via \Coq's tactics.

\section{Branch prediction}\label{sec:branch}

In computer architecture, a \emph{branch predictor} is a construct that tries to guess which  branch the control will exit, \eg\ in an {\tt if-then-else}, before the result of the test is actually known. Such a construct is convenient when the evaluation of the test is so much more time demanding \wrt\ executing the other instructions, that the time lost, when having to backtrack because the guess was wrong, is significantly compensated by the speed-up which is achieved, when the guess is correct.

In this section we model the behavior of such a structure in \LLFP\ in the case of the \emph{Unlimited Register Machine (URM)}, a simple universal model of computation popularized by Cutland \cite{Cutland:computability}.

An URM has an infinite number of registers $R_0, R_1, \ldots$ containing natural numbers $r_0, r_1, \ldots$ which may be mutated by instructions. Sequences of instructions form programs:
\[
    \begin{array}{lclll}
    s & ::= & \langle \iota {\mapsto} r_{\iota} \rangle^\rangeiNinf &
    & \quad \textrm{Store}
\\
    I & ::= & Z(i) \mid S(i) \mid T(i,j) \mid J(i,j,k) & \quad i,j,k {\in} \mathbb{N}
    & \quad \textrm{Instruction}
    \\
    P & ::= & (\iota {\mapsto} I_\iota)^\rangeiN\ & \quad m {\in} \mathbb{N}
    & \quad \textrm{Program}
\\
    \end{array}
\]
The four kinds of instructions Zero, Successor, Transfer, Jump have the following intended meanings ($r \loads R$ stands for loading the natural value $r$ into the register $R$):
\[
    \begin{array}{lcl}
    Z(i) & \triangleq & 0 \loads R_i
\\
    S(i) & \triangleq & r_i + 1 \loads R_i
\\
    T(i,j) & \triangleq & r_i \loads R_j
\\
    J(i,j,k) & \triangleq & \textrm{if $r_i {=} r_j$ then execute as next instruction the $k$-th instruction else the next one}
    \end{array}
\]
When given a program $P$, a program counter $n$, and a store $s$, an URM executes the program starting from the $n$-th instruction in $P$ and carries out the instructions sequentially (unless a positive {\em J} instruction is encountered),  mutating at each step the contents of the store as prescribed by the instructions.
The evaluation of a program may be described therefore, as follows:
\[
\begin{array}{lcl}
E(P,n,s) & = &
\left\{ \begin{array}{ll}
	s & \quad \textrm{if $fetch(P,n){=}Halt$}
\\
	E(P,n{+}1,zero(s,i)) & \quad \textrm{if $fetch(P,n){=}Z(i)$}
\\
	\ldots & \quad \ldots
\\
	E(P,k,s) & \quad \textrm{if $fetch(P,n){=}J(i,j,k)$ and $s(i){=}s(j)$}
\\
	E(P,n{+}1,s) & \quad \textrm{if $fetch(P,n){=}J(i,j,k)$ and $s(i){\neq}s(j)$}
	\end{array}
\right.
\end{array}
\]
We use the $zero$ function for updating the store according to the {\em Z} instruction (similar updating functions $succ$ for {\em S} and $move$ for {\em T} are omitted) and the $fetch$ function for recovering the instruction pointed to by the program counter. The $Halt$ instruction is added to make the function $fetch$ total. A computation stops if and only if $fetch$, fetches $Halt$. On the other hand, due to the looping back via the $J$ instruction, there are non-terminating computations. In our case study we consider only terminating computations (the interested reader may refer to \cite{alberto:urm} for a coinductive approach to diverging computations).

The functions introduced in order to formalize evaluation are defined as follows:
\[
   \begin{array}{lcl}
fetch(P,n) & \triangleq & \textrm{if $n {>} length(P)$ then $Halt$ else $I_n$}
\\
zero(s,i) & \triangleq & \lambda \iota {\in} \mathbb{N}.\; \textrm{if $\iota {=} i$ then $0$ else $s(\iota)$}
\\
succ(s,i) & \triangleq & \lambda \iota {\in} \mathbb{N}.\; \textrm{if $\iota {=} i$ then $s(\iota){+}1$ else $s(\iota)$}
\\
move(s,i,j) & \triangleq & \lambda \iota {\in} \mathbb{N}.\; \textrm{if $\iota {=} j$ then $s(i)$ else $s(\iota)$}
\\
   \end{array}
\]

To introduce an \LLFP\ signature, for the URM machine, we need first to encode infinite stores and non-structured programs. Both datatypes are handled by mimicking lists.

\begin{definition}[\LLFP\ signature for Stores and Programs]\label{branch-syntax}
\[
\begin{array}{ll}
\verb|nat: Type| &
\verb|0: nat| \qquad \verb|S: nat| \to \verb|nat|
\\
\verb|store: Type| &
\verb|zeros: store| \qquad
\verb|cs: nat| \to \verb|store| \to \verb|store|
\\
\verb|ins: Type| &
\verb|Ht: ins| \qquad \verb|Zr: nat| \to \verb|ins| \qquad \ldots \qquad
\verb|Jp: nat| \to \verb|nat| \to \verb|nat| \to \verb|ins|
\\
\verb|pgm: Type| &
\verb|void: pgm| \qquad
\verb|cp: ins| \to \verb|pgm| \to \verb|pgm|
\end{array}
\]
\end{definition}
Natural numbers \texttt{nat} are extensively used in the URM-signature: actually, we make them play also the role of store locations, \eg\ in \texttt{Zr} (encoding {\em Z}), and program counters, in \texttt{Jp} (encoding {\em J}).
As far as stores, we use the nil-like \texttt{zeros} constructor which represents an infinite sequence of {\tt 0} values. Stores may be updated on demand via the cons-like \texttt{cs} constructor.
We encode programs, similarly, as lists of instructions in \texttt{ins}, with the addition of \texttt{Ht}, which represents the $Halt$ instruction  motivated above.

To structure the evaluation of URM programs, we introduce the two small-step relations $\leadsto \;\subseteq\; pgm \times nat \times store \times nat \times store$ and $\Rightarrow \;\subseteq\; pgm \times nat \times store \times store$, as follows:
\[
   \begin{array}{c}
     \infer[\textrm{(eZ)}]
     {\langle n, s \rangle \leadsto^P \langle n{+}1, zero(s,i) \rangle}
     {fetch(P,n) {=} Z(i)}
\qquad
\infer[\textrm{(eS)}]
     {\langle n, s \rangle \leadsto^P \langle n{+}1, succ(s,i) \rangle}
     {fetch(P,n) {=} S(i)}
\\
\\
     \infer[\textrm{(eT)}]
     {\langle n, s \rangle \leadsto^P \langle n{+}1, move(s,i,j) \rangle}
     {fetch(P,n) {=} T(i,j)}
\qquad
     \infer[\textrm{(trans)}]
     {\langle n, s \rangle \leadsto^P \langle q, u \rangle}
     {\langle n, s \rangle \leadsto^P \langle m, t \rangle \quad
	  \langle m, t \rangle \leadsto^P \langle q, u \rangle}
\\
\\
     \infer[\textrm{(Jt)}]
     {\langle n, s \rangle \leadsto^P \langle k, s \rangle}
     {fetch(P,n) {=} J(i,j,k) \quad s(i){=}s(j)}
\qquad
     \infer[\textrm{(Jf)}]
     {\langle n, s \rangle \leadsto^P \langle n{+}1, s \rangle}
     {fetch(P,n) {=} J(i,j,k) \quad s(i){\neq}s(j)}
\\
\\
     \infer[\textrm{(empty)}]
     {\langle n, s \rangle \Rightarrow^P s}
     {fetch(P,n) {=} halt}
\qquad
     \infer[\textrm{(stop)}]
     {\langle n, s \rangle \Rightarrow^P t}
     {\langle n, s \rangle \leadsto^P \langle m, t \rangle \quad fetch(P,m) {=} halt}
\end{array}
\]
Now we come to the crucial issue. \LLFP's \emph{lock-types} allow us to model faithfully also the execution of a ``branch prediction'' version of this semantics, by postponing the double access to the store and test  required by $J$, which is a slow instruction.
Lock-types permit to carry out the double access and equality check concurrently and asynchronously \wrt\ the main computation, in the spirit of the  ``fast and loose'' philosophy. We omit for simplicity in the following definition the encoding of the {\em S} and {\em T} instructions.

\begin{definition}[\LLFP\ signature for Evaluation]\label{branch-semantics}
\[
\begin{array}{lcl}
\verb|T| & : & \verb|Type|
\\
\verb|fetch| & : & \verb|pgm| \to \verb|nat| \to \verb|ins| \to \verb|Type|
\\
\verb|zero| & : & \verb|store| \to \verb|nat| \to \verb|store| \to \verb|Type|
\\
\verb|step| & \verb|:| & \verb|prg| \to \verb|nat| \to \verb|store| \to \verb|nat| \to \verb|store| \to \verb|Type|
\\
\verb|eval| & \verb|:| & \verb|prg| \to \verb|nat| \to \verb|store| \to \verb|store| \to \verb|Type|
\\
\mathtt{\langle \_,\_,\_ \rangle} & : & \verb|store| \to \verb|nat| \to \verb|nat| \to \verb|T|
\\
\verb|fvn| & : & \Pi \verb|n:nat|.\;
\verb|fetch void n Ht|
\\
\verb|fc0| & : & \Pi \verb|I:ins|.\; \Pi \verb|Q:prg|.\;
\verb|fetch (cp I Q) 0 I|
\\
\verb|fcn| & : & \Pi \verb|I,L:ins|.\; \Pi \verb|Q:prg|.\; \Pi \verb|n:nat|.\;
\verb|fetch Q n L| \to \verb|fetch (cp I Q) (S n) L|
\\
\verb|zvn| & : & \Pi \verb|n:nat|.\;
\verb|zero zeros n zeros|
\\
\verb|zc0| & : & \Pi \verb|v:nat|.\; \Pi \verb|s:store|.\;
\verb|zero (cs v s) 0 (cs 0 s)|
\\
\verb|zcn| & : & \Pi \verb|v,n:nat|.\; \Pi \verb|s,t:store|.\;
\verb|zero s n t| \to \verb|zero (cs v s) (S n) (cs v t)|
\\
\verb|sZ| & \verb|:| & \Pi \verb|P:pgm|.\; \Pi \verb|n,i:nat|.\;
\Pi \verb|s,t:store|.\\
              &            &
\verb|fetch P n (Z i)| \to \verb|zero s i t| \to \verb|step P n s (S n) t|
\\
\end{array}
\]
\[
\begin{array}{lcl}
\verb|sJt| & \verb|:| & \Pi \verb|P:pgm|.\; \Pi \verb|n,i,j,k:nat|.\; \Pi \verb|s:store|.\\
              &            & \verb|fetch P n (J i j k)| \to
                               \Lock{\scriptsize\tt{Eq}}
								{\scriptsize{\mathtt{\langle{s},{i},{j}\rangle}}}	
								{\scriptsize\mathtt{{T}}}
								{\mathtt{step\ \,P\ \, n\ \,s\ \,k\ \,s}}
\\
\verb|sJf| & \verb|:| & \Pi \verb|P:pgm|.\; \Pi \verb|n,i,j,k:nat|.\; \Pi \verb|s:store|.\\
              &            & \verb|fetch P n (J i j k)| \to
                               \Lock{\scriptsize\tt{Neq}}
								{\scriptsize{\mathtt{\langle{s},{i},{j}\rangle}}}	
								{\scriptsize\mathtt{{T}}}
								{\mathtt{step\ \,P\ \, n\ \,s\ \,(S\ \, n)\ \,s}}
\\
\verb|sTr| & \verb|:| & \Pi \verb|P:pgm|.\; \Pi \verb|n,m,q:nat|.\; \Pi \verb|s,t,u:store|.\\
              &            & \verb|step P n s m t| \to \verb|step P m t q u| \to \verb|step P n s q u|
\\
\verb|e0| & \verb|:| & \Pi \verb|P:pgm|.\; \Pi \verb|n:nat|.\; \Pi \verb|s:store|.\;
\verb|fetch P n halt| \to \verb|eval P n s s|
\\
\verb|e1| & \verb|:| & \Pi \verb|P:pgm|.\; \Pi \verb|n,m:nat|.\; \Pi \verb|s,t:store|.\\
              &            & \verb|step P n s m t| \to \verb|fetch P m halt| \to \verb|eval P n s t|
\\
\end{array}
\]
where $\mathtt{Eq}(\Gamma \VDASHS \mathtt{\langle s,i,j \rangle : T})$ holds iff $s(i) {=} s(j)$, and $\mathtt{Neq}(\Gamma \VDASHS \mathtt{\langle s,i,j \rangle : T})$ iff $s(i) {\neq} s(j)$.
\end{definition}

We now handle this second case study via the \Coq\ editor introduced in Section \ref{sec:implementation}.
We take advantage of built-in natural numbers and lists to define stores, instructions, and programs (all typed by {\tt Set}), namely:
\begin{verbatim}
Definition store: Set := list nat.
Parameter ins: Set. Parameter Ht: ins. ...
Definition pgm: Set := list ins.
\end{verbatim}
The input to the oracle, \ie\ a store and a pair of locations, is defined  as an inductive type {\tt T} of triples and corresponding projection functions.
Memory access is realized through the built-in total function {\tt nth},
which returns the {\tt 0} value when the end of a list-store is reached.
The oracle predicates can then be formalized in \Coq\ by using these datatypes, as follows:
\begin{verbatim}
Inductive T: Set := triple: store -> nat -> nat -> T.
Definition pr1 (x:T): store := match x with triple s i j => s end. ...
Definition s_nth (s:store) (n:nat): nat := nth n s 0.
Definition Eq := fun x:T => s_nth (pr1 x) (pr2 x) = s_nth (pr1 x) (pr3 x).
\end{verbatim}
The evaluation semantics is finally encoded as a predicate, via suitable auxiliary functions that update the store (we omit for lack of space such functions and most of the \Coq\ translation of Definition~\ref{branch-semantics}, it is available at \cite{web-appendix}):
\begin{verbatim}
Parameter step: pgm -> nat -> store -> nat -> store -> Prop.
Parameter sJt: forall P n i j k s, fetch P n = (Jp i j k) ->
               lockF T (triple s i j) Eq (step P n s k s). ...
Parameter eval: pgm -> nat -> store -> store -> Prop. ...
\end{verbatim}

In order to appreciate the encoding at work, let us consider the simple  program $P \triangleq Z(0), J(0,1,0)$ and the stores $s \triangleq 1{:}1{:}zeros$ and $t \triangleq 0{:}1{:}zeros$.
Then we have the fragment derivation\footnote{In the present and the next derivations we display \LLFP's types without the proof terms because these are synthesized by the editor.}:
\[
\infer[\textrm{(sTr)}]
		{\langle 1, s \rangle \leadsto^P \langle 1, t \rangle
		}
		{
		\infer[\textrm{(O${\cdot}$Top)}]
		{\langle 1, s \rangle \leadsto^P \langle 0, s \rangle}
		{\infer[\textrm{(sJt)}]
			{
			\Lock{Eq}
				{\langle s,0,1 \rangle}{T}
				{\langle 1, s \rangle \leadsto^P \langle 0, s \rangle}
			}
			{P(1) {=} J(0,1,0)
			} & Eq (\langle s,0,1 \rangle)
		}
         &
		\infer[\textrm{(sZ)}]
			{\langle 0, s \rangle \leadsto^P \langle 1, t \rangle
			}
			{P(0) {=} Z(0)
			}				
		}
\]
In this proof tree there is a limited amount of parallelism, because we wait until the verification of $Eq (\langle s,0,1 \rangle)$ is accomplished, before channeling the reductions via the transitivity $(sTr)$ rule.
The parallelism may be increased by exploiting the $(O{\cdot}Guarded{\cdot}Unlock)$ rule, which handles arguments within a lock-type, and allows us to apply the $(sTr)$ rule even in the presence of a left-hand $J$ reduction:
\[
\infer[\textrm{(O${\cdot}$Top)}]
{\langle 1, s \rangle \leadsto^P \langle 1, t \rangle
}
{
\infer[\textrm{(sTr via O${\cdot}$Guarded${\cdot}$Unlock)}]
		{
		\Lock{Eq}{\langle s,0,1 \rangle}{T}
			{\langle 1, s \rangle \leadsto^P \langle 1, t \rangle}
		}
		{\infer[]
			{
			\Lock{Eq}
				{\langle s,0,1 \rangle}{T}
				{\langle 1, s \rangle \leadsto^P \langle 0, s \rangle}
			}
			{P(1) {=} J(0,1,0)
			}
         &
		\infer[]
			{\langle 0, s \rangle \leadsto^P \langle 1, t \rangle
			}
			{P(0) {=} Z(0)
			}				
		}
&
Eq (\langle s,0,1 \rangle)
}
\]
The $Eq (\langle s,0,1 \rangle)$ check can now be delayed, and carried out independently \wrt\ the main reduction.
The $(O{\cdot}Guarded{\cdot}Unlock)$ rule allows for more proof trees for the same judgment. This is precisely what accommodates the ``branch prediction'' philosophy.

An even higher degree of parallelism could be achieved in \LLFP\ if a mechanism to ``compose" pieces of reductions within \emph{different} lock-types were available. This would give us the opportunity to apply the transitivity rule ``under'' \emph{pairs} of Jump instructions.
If, for instance, we want to manage a maximum of $2$ branch predictions, we can define introduction and elimination rules of the following shape:
\[
\begin{array}{c}
\infer[(\P_+)]
{\Lock{\P_1;\P_2}
    {\langle \vec{x_1} \rangle; \langle \vec{x_2} \rangle}{T}
    {\langle n, s \rangle \leadsto^P \langle q, u \rangle}}
{\Lock{\P_1}
   {\langle \vec{x_1} \rangle}{T}
   {\langle n, s \rangle \leadsto^P \langle m, t \rangle}
&
\Lock{\P_2}
    {\langle \vec{x_2} \rangle}{T}
    {\langle m, t \rangle \leadsto^P \langle q, u \rangle}
}
\\
\\
\infer[(\P_{-1})]
{\Lock{\P_2}
   {\langle \vec{x_2} \rangle}{T}
   {\langle n, s \rangle \leadsto^P \langle m, t \rangle}}
{\Lock{\P_1;\P_2}
    {\langle \vec{x_1} \rangle; \langle \vec{x_2} \rangle}{T}
    {\langle n, s \rangle \leadsto^P \langle m, t \rangle}
\quad \P_1 (\vec{x_1})
}
\quad
\infer[(\P_{-2})]
{\Lock{\P_1}
   {\langle \vec{x_1} \rangle}{T}
   {\langle n, s \rangle \leadsto^P \langle m, t \rangle}}
{\Lock{\P_1;\P_2}
    {\langle \vec{x_1} \rangle; \langle \vec{x_2} \rangle}{T}
    {\langle n, s \rangle \leadsto^P \langle m, t \rangle}
\quad \P_2 (\vec{x_2})
}
\end{array}
\]
where $\P_\iota$ stands for $Eq$ or $Neq$, $\vec{x_\iota} \equiv \langle x_\iota, i_\iota,j_\iota \rangle$, and
$\P_\iota (\vec{x_\iota}) \equiv Eq(\langle x_\iota,i_\iota,j_\iota \rangle)$ if $\P_\iota \equiv Eq$ or $\P_\iota (\vec{x_\iota}) \equiv Neq(\langle x_\iota,i_\iota,j_\iota \rangle)$ if $\P_\iota \equiv Neq$, for all $\iota {\in} \{1,2\}$.
We could then delay even more the access to pairs of memory locations for checking for (dis)equality of their contents:
\[
\infer[]
{
\langle 1, s \rangle \leadsto^P \langle 2, t \rangle
}
{
\infer[(\P_{-1})]
{
\Lock{Neq}
  {\langle t,0,1 \rangle}{T}
  {\langle 1, s \rangle \leadsto^P \langle 2, t \rangle}
}
{
\infer[(\P_+)]
	{
	\Lock{Eq;Neq}
		{\langle s,0,1 \rangle; \langle t,0,1 \rangle}{T}
		{\langle 1, s \rangle \leadsto^P \langle 2, t \rangle}
	}
	{\infer[]
		{
		\Lock{Eq}{\langle s,0,1 \rangle}{T}
			{\langle 1, s \rangle \leadsto^P \langle 1, t \rangle}
		}
		{\vdots			
		}
&
\infer[]
{\Lock{Neq}
  {\langle t,0,1 \rangle}{T}
  {\langle 1, t \rangle \leadsto^P \langle 2, t \rangle}
}
{P(1) {=} J(0,1,0)
}
	}
&
Eq(\langle s,0,1 \rangle)
}
&
Neq(\langle t,0,1 \rangle)
}
\]
We will focus on these envisaged extensions and corresponding encodings in the following Section \ref{sec:LLFP-ext}.
We anticipate here that the ``composition" of predicates can be dealt with via lock nesting, that is, we manage elimination rules in the form $\P_-$ by means of the $(O{\cdot}Top{\cdot}Unlock)$ rule (\ie\ \Coq's \texttt{top_unlock} lemma), and we manage introduction rules such as $\P_+$ by ``\texttt{unfold}"ing the \texttt{lockF} constructor.

In conclusion, in this section, we have shown how \LLFP\ can naturally accommodate computations running in parallel asynchronously, as it happens when performing branch prediction.

\section{Towards an algebra of locks}\label{sec:LLFP-ext}

In the previous section we have informally argued about possible extensions of \LLFP\ in order to accommodate logical combinations of predicates in locks. 
In fact, the branch prediction case study has pointed out on the one hand the need of ``conjunctions" of lock predicates (in order to augment the parallelism of execution), on the other hand the possibility of managing ``disjunctions" of lock predicates (to represent in a compact way pairs of mutually exclusive computations).
Therefore, we would like to handle both conjunctions and disjunctions of lock predicates, according to the following introduction rules:
\[
  \infer[(O{\cdot}Lock{\cdot}{\wedge})]
  {\Gamma \VDASHS {\Lock {\P_1\wedge\P_2} {\langle N_1,N_2\rangle} {\langle\sigma_1,\sigma_2\rangle} M} : {\Lock {\P_1\wedge\P_2} {\langle N_1,N_2\rangle} {\langle\sigma_1,\sigma_2\rangle} {\rho}}}
  {\Gamma \VDASHS {\Lock {\P_1} {N_1} {\sigma_1} M} : {\Lock {\P_1} {N_1} {\sigma_1} {\rho}}\qquad
   \Gamma \VDASHS {\Lock {\P_2} {N_2} {\sigma_2} M} : {\Lock {\P_2} {N_2} {\sigma_2} {\rho}}
  }
\]
\[
  \infer[(O{\cdot}Lock{\cdot}{\oplus})]
  {\Gamma \VDASHS {\Lock {\P_1\oplus\P_2} {N} {\sigma} {[M_1,M_2]}} : {\Lock {\P_1\oplus\P_2} {N} {\sigma} {\rho_1\oplus\rho_2}}}
  {{\Gamma \VDASHS M_1 : \rho_1}\qquad {\Gamma \VDASHS M_2 : \rho_2}
  }
\]
where $[M_1,M_2]$ denotes the ``bookkeeping'' of the terms $M_1$ and $M_2$ of types $\rho_1$ and $\rho_2$, respectively, into a new \emph{binary record} structure. Indeed, $\rho_1\oplus\rho_2$ represents the record type whose components are of types $\rho_1$ and $\rho_2$, and $\P_1$ and $\P_2$ are two {\em mutually exclusive} predicates.
The $\oplus$ type is eliminated as follows:
\[
  \infer[(O{\cdot}Lock{\cdot}{\oplus_l})]
   {\Gamma \VDASHS ({\Unlock {\P_1\oplus\P_2} N \sigma M})_l : (\rho)_l}
   {\Gamma \VDASHS M : {\Lock {\P_1\oplus\P_2} N \sigma \rho}  \qquad {\P_1}({\Gamma\VDASHS N:\sigma})
\qquad\P_1\ \textrm{and}\ \P_2\ \textrm{are mutually exclusive}
}
\]
\[
  \infer[(O{\cdot}Lock{\cdot}{\oplus_r})]
   {\Gamma \VDASHS ({\Unlock {\P_1\oplus\P_2} N \sigma M})_r : (\rho)_r}
   {\Gamma \VDASHS M : {\Lock {\P_1\oplus\P_2} N \sigma \rho}  \qquad {\P_2}({\Gamma\VDASHS N:\sigma})
\qquad\P_1\ \textrm{and}\ \P_2\ \textrm{are mutually exclusive}}
\]
where $(M)_l$, respectively $(M)_r$, represents the left, respectively right, component of the binary record term $M$, and $(\rho)_l$, respectively $(\rho)_r$, represents the left, respectively right, component of the binary record type $\rho$. Due to lack of space, we omit the obvious elimination rules for the conjunction of lock predicates, and their nested equivalents.

Given our shallow encoding of \LLFP\ in \Coq, such derived rules can be rendered very easily introducing two new definitions:
\begin{verbatim}
Definition lockF_and :=
  fun s1: Set => fun N1: s1 => fun P1: s1 -> Prop =>
  fun s2: Set => fun N2: s2 => fun P2: s2 -> Prop =>
  fun r: Prop => forall x: P1 N1, forall y: P2 N2, r.

Definition lockF_xor :=
  fun s: Set => fun N: s => fun P1: s -> Prop => fun P2: s -> Prop =>
  fun r1 r2: Prop => xor (P1 N) (P2 N) ->
  xor (forall x: P1 N, r1) (forall y: P2 N, r2).
\end{verbatim}
where mutual exclusion is encoded as follows:
\begin{verbatim}
Definition xor := fun A B:Prop => (A /\ not B) \/ (not A /\ B).
\end{verbatim}
So doing, we can formally prove the following lemmata:
\begin{verbatim}
Lemma lock_and: forall s1: Set, forall N1: s1, forall P1: s1 -> Prop,
  forall s2: Set, forall N2: s2, forall P2: s2 -> Prop,
  forall r: Prop, forall x: P1 N1, forall y: P2 N2,
  lockF_and s1 N1 P1 s2 N2 P2 r <-> lockF s1 N1 P1 (lockF s2 N2 P2 r).

Lemma lock_xor: forall s: Set, forall N: s, 
  forall P1: s -> Prop, forall P2: s -> Prop,
  forall r1 r2: Prop, forall x: P1 N, forall y: P2 N,
  xor (P1 N) (P2 N) ->
  (lockF_xor s N P1 P2 r1 r2 <-> xor (lockF s N P1 r1) (lockF s N P2 r2)).
\end{verbatim}
In other words, \texttt{lockF\_and} is syntactic sugar for lock nesting, while \texttt{lockF\_xor} reduces to an exclusive disjunction between two \texttt{lockF} judgments.

We remark that alternative approaches to the development of an algebra of locks may be pursued.

\section{Conclusion}\label{sec:future}

This paper provides two contributions to the development of the Lax Logical Framework \LLFP, introduced in \cite{lmcs:3771}.
The first contribution is a very ``shallow'', actually definitional, implementation of \LLFP\ in \Coq. This produces immediately a proof development environment, supporting mechanized proof search, for a version of \LLFP\ in which the predicates used in locks are \Coq-definable.
The second contribution shows how the feature of \LLFP, which allows for postponing the evaluation of an ultimately proof-irrelevant side-condition, can model naturally instances of the emerging paradigm of ``fast and loose'' reasoning, \cite{FL}. Actually, we can say that the philosophy of locks amounts to applying such a paradigm at a metatheoretic level.
Both contributions are essential in the development of the case study reported in the paper concerning {\em branch prediction}, which is a form of ``fast and loose'' evaluation, of the URM machine. 

We do not provide a formal adequacy theorem for the branch prediction case study. We are currently working on it as well as other ``fast and loose'' reasoning patterns. This is problematic however, since they are not fully spelled out in the literature. We believe that adequacy would be very significant because it would provide a thorough understanding of the heuristics underpinning such paradigms. 

Both contributions appear to be rather fruitful. The definitional implementation suggests how to rapidly prototype editors for other calculi such as \CLLFPQ, see \cite{lmcs:3771}, or extensions of \LLFP\ which support an algebraic structure of locks as outlined in Section \ref{sec:LLFP-ext}. 

More case studies need to be developed. For lack of space, we could not even outline here another seminal case-study, namely that on {\em optimistic concurrency control}, which is another important example of the ``fast and loose'' paradigm applied to non-interference issues.
Another important case study related to the ``fast and loose'' philosophy which we intend to develop is that of Fitch-Prawitz consistent Set Theory, \cite{APLAS}. This is the natural counterpart of the na\"{\i}ve Set Theory used in developing ordinary mathematics.

It would be interesting to address the issue of extending  full-fledged locks to \Coq\ itself. 

Finally, we intend to explore how to prototype an alternate editor for \LLFP\ using the MMT UniFormal Framework of F.~Rabe, \cite{DBLP:journals/iandc/RabeK13}.

\nocite{*}
\bibliographystyle{eptcs}
\bibliography{biblio}

\end{document}